# Numerical Assessment and Optimization of Discrete-Variable Time-Frequency Quantum Key Distribution

*Jasper Rödiger[1,2], Nicolas Perlot[1], Roberto Mottola[1,2], Robert Elschner[1], Carl-Michael Weinert[1], Oliver Benson[2], Ronald Freund[1]*

[1]*Fraunhofer Heinrich Hertz Institute, Einsteinufer 37, 10587 Berlin, Germany*
[2]*Humboldt-Universität zu Berlin, AG Nanooptik, Newtonstraße 15, 12489 Berlin, Germany*

The discrete variables (DV) time-frequency (TF) quantum key distribution (QKD) protocol is a BB84 like protocol, which utilizes time and frequency as complementary bases. As orthogonal modulations, pulse position modulation (PPM) and frequency shift keying (FSK) are capable of transmitting several bits per symbol, i.e. per photon. However, unlike traditional binary polarization shift keying, PPM and FSK do not allow perfectly complementary bases. So information is not completely deleted when the wrong-basis filters are applied. Since a general security proof does not yet exist, we numerically assess DV-TF-QKD. We show that the secret key rate increases with a higher number of symbols per basis. Further we identify the optimal pulse relations in the two bases in terms of key rate and resistance against eavesdropping attacks.

## 1 Introduction

Quantum key distribution (QKD), the first applicable quantum technology, is used to distribute a secret key to two parties, which can then for example be used as a one-time pad for absolute secure communication. In the conventional implementation of the BB84 protocol for QKD either polarization or phase of single photons is used to form the two required non-orthogonal bases [1-3]. Yet, in various scenarios modulation of polarization or phase may not be practicable. For example for satellite communications polarization is usually set to left- or right-handed circular [4, 5]. Here we investigate a BB84-like protocol where information is coded either via time or via frequency modulations. More specifically pulse position modulation (PPM) and frequency shift keying (FSK) are used to form the two required complementary bases. This protocol, namely the time-frequency (TF)-QKD protocol is promising, because PPM and FSK rely on techniques in classical communication, which makes this protocol easier to implement in existing systems and networks.

The TF-QKD protocol was suggested by several authors [6, 7]. The security of embodiments of continuous variable (CV)-TF-QKD was addressed [8-12] while a security proof against general attacks for discrete variables (DV) is not yet at hand. DV implementations using prepare-and-measure [13] or entanglement [14] were reported. Recently DV-TF-QKD was addressed with respect to a certain intercept/resend attack [15] and turbulent free-space channels [16].

The main conceptual difference between DV-TF-QKD and BB84 is the imperfect complementarity of the measurement bases of DV-TF-QKD. In BB84, getting all information coded in one basis completely deletes the information coded in the other. However, in DV-TF-QKD not all information is deleted.

In [15] the performance of prepare-and-measure DV-TF-QKD is investigated. There, narrow pulses compared to bin width and pulse separation are assumed. Using narrow pulses was proposed in [12] for CV-, not DV-TF-QKD. In [15] the protocol performance was calculated based on a variation of the intercept/resend attack where the gaps between the pulses are exploited by an eavesdropper to distinguish the bases. In CV-TF-QKD narrow pulses are preferable since there are no gaps and both bases can be made indistinguishable [12]. However, this is not the case for DV-TF-QKD, thus narrow pulses might not be the optimal choice here.

In this paper we show that a larger overlap of the pulses although leading to more errors in the raw key can be preferable over narrow pulses. After introducing the basics of the protocol we will investigate how the widths of the symbol-pulses affect the secret key rate of DV-TF-QKD. We will also find the optimal pulse width of the conjugated pulses. Further, we will calculate the number of secret bits per photon for the optimal pulse width and show that a higher number of possible symbols per basis increases the secret key rate regarding an intercept/resend attack exploiting the unique probabilities of DV-TF-QKD.

## 2 The time-frequency protocol

In DV-TF-QKD like in BB84 single photons coded in two bases are used to distribute a secret key between sender Alice and receiver Bob hidden from an eavesdropper Eve. Alice and Bob choose their sending/measurement basis randomly and afterwards discard all measurement results where their bases differ. This is called sifting. Then they compare a small fraction of the resulting sifted key which they discard afterwards to measure the quantum symbol error rate (QSER) and conclude on Eve's knowledge of the key. If Eve's knowledge is low enough, it is possible to distill a secret key using error correction and privacy amplification, otherwise the QKD process needs to be repeated.

### 2.1 Pulse relationships

The modulations used as the two bases are $M$-PPM and $M$-FSK as shown in Fig. 1. Here $M$ symbol pulses in $M$ different time (resp. frequency) bins represent $M$ different symbols. In the conjugate basis, i.e. frequency for time pulses and vice versa, the conjugate pulses spread out over all bins and thus contain no information. Because each pulse contains exactly one photon, the pulse energy densities represent the probability distributions to measure the photon in a certain point in time (resp. frequency). Without loss of generality we assume Fourier-limited Gaussian pulses. Each symbol can contain up to $N = \log_2(M)$ raw bits, i.e. one qu-$N$-it. In Fig. 1 shows the energy densities for $M = 4$.

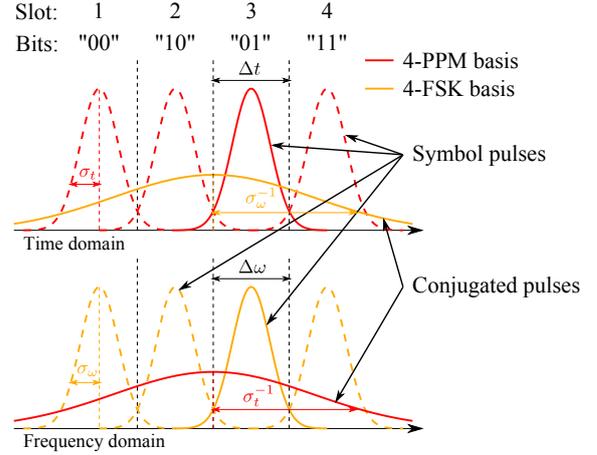

**Fig. 1: (Color online) PPM- and FSK-symbol and conjugated pulses in the time and frequency domain for $M = 4$.**

$\sigma_t$ (resp. $\sigma_\omega$) is half the $1/e$ width of the energy density of a PPM (resp. FSK) symbol pulse and $\Delta t$ (resp. $\Delta \omega$) is the symbol pulse separation. The central frequency (resp. time) of the conjugated PPM (resp. FSK) pulses is chosen such that they are centered with respect to the FSK (resp. PPM) symbol pulses in the frequency (resp. time) domain. The $1/e$ widths of the symbol pulses and conjugate pulses are reciprocal. The pulse energy density is given by

$$\rho_\sigma(z) \equiv |\psi_\sigma(z)|^2 \equiv \frac{\sqrt{2}}{\sigma} \phi\left(\sqrt{2}\frac{z}{\sigma}\right). \qquad (1)$$

Here $\phi(z) = 1/(2\pi)^{1/2} \exp\{-z^2/2\}$ is the standard normal distribution, $\sigma$ is equal to $\sigma_t$ (resp. $\sigma_\omega$) for the PPM (resp. FSK) symbol pulse and $\sigma_\omega^{-1}$ (resp. $\sigma_t^{-1}$) for the conjugated FSK (resp. PPM) pulse. Furthermore $z = t$ for time (resp. $z = \omega$ for frequency) pulses.

In BB84 it is impossible to reveal any information of the basis by a measurement. As pointed out in [12] in CV-TF-QKD narrow symbol pulses are preferable because hereby it is possible to make the symbol pulses indistinguishable from a conjugated pulses of the complementary basis. This means in the frequency (resp. time) domain all FSK (resp. PPM) symbol pulses can perfectly overlap with all conjugate PPM (resp. FSK) pulses when the symbol pulses rate is varied over the frequency (resp. time) accordingly. In DV-TF-QKD this is not possible for both bases. However, with optimized pulse widths of symbol and conjugated pulses the overlap can be improved significantly.

Additionally for BB84 the measurement in the wrong basis deletes all information in the other (correct) basis. However, in the TF-protocol it is possible to extract information on the encoded qu-$N$-it even if it is already filtered in the wrong basis. For example Eve could accomplish this by guiding the photons through two successive time and frequency filters. To hinder this approach the first filtering process should make the outcome of the second perfectly random. That means its time distribution should be significantly broadened by projection on any FSK symbol. In DV-TF-QKD this condition cannot be achieved perfectly.

Both, perfect overlap and information-deletion, can be approached by applying the following pulse relations:

i. $\Delta t \approx 2\sigma_t$ : the $1/e$ width of the PPM symbol pulses is similar to the pulse distance,
ii. $\Delta\omega \approx 2\sigma_\omega$ : the $1/e$ width of the narrow FSK symbol pulses is similar to the pulse distance,
iii. $2\sigma_t^{-1} \approx M\Delta\omega$ : the conjugated PPM pulse is approximately as wide as the $M$ FSK symbol pulses,
iv. $2\sigma_\omega^{-1} \approx M\Delta t$ : the conjugated FSK pulse is approximately as wide as the $M$ PPM symbol pulses.

To extend these approximations to quantitative equalities we define the normalized pulse widths

$$\alpha \equiv \frac{2\sigma_t}{\Delta t} = \frac{2\sigma_\omega}{\Delta\omega} \qquad (2)$$

and

$$\beta \equiv \frac{2\sigma_\omega^{-1}}{M\Delta t} = \frac{2\sigma_t^{-1}}{M\Delta\omega}, \qquad (3)$$

expecting the values for both parameters to be in the order of one (analogue to i. to iv.). $\alpha$ is half the normalized $1/e$ width of the amplitude $\Psi_\sigma(z)$ of the symbol pulses and $\beta$ of the conjugated pulses. For simplicity we assume the pulse relations to be the same for PPM and FSK pulses with the result that time and frequency will be interchangeable in the following calculations. Further, we consider perfect single photon sources and do not consider any noise so that each pulse is undistorted and contains exactly one photon. Because there is only one photon per pulse, there is no ambiguity in the determination of the received symbol. Classical PPM systems generally have dead times between each symbol [17], thus we neglect inter-symbol interference.

It is convenient to define the pulse and bin positions at this stage. With $i = [1, M]$ the centers of PPM (resp. FSK) symbol pulses, normalized to $\Delta t$ (resp. $\Delta\omega$), are defined as

$$c(i) = i - \frac{M+1}{2}. \qquad (4)$$

We assume rectangular bins for Bob and Eve, since it is easy to calculate, but also since it prevents additional errors. Each bin is defined by its lower $b_{\text{low}}(j)$ and upper bound $b_{\text{up}}(j)$ which we normalize to $\Delta t$ (resp. $\Delta\omega$) for time (resp. frequency):

$$b_{\text{low}}(j) = \begin{cases} -\infty & \text{for } j = 1 \\ j - \tfrac{1}{2}M - 1 & \text{for } j = 2,...,M \end{cases} \qquad (5)$$

$$b_{\text{up}}(j) = \begin{cases} j - \tfrac{1}{2}M & \text{for } j = 1,...,M-1 \\ +\infty & \text{for } j = M \end{cases} \qquad (6)$$

Outer bins (i.e. $j = 1$ or $M$) are considered infinitely wide in order to maintain all photons. Fig. 2 shows the bin and symbol pulse positions for $M = 4$. The part of the pulse which spills in neighboring bins is called spill region.

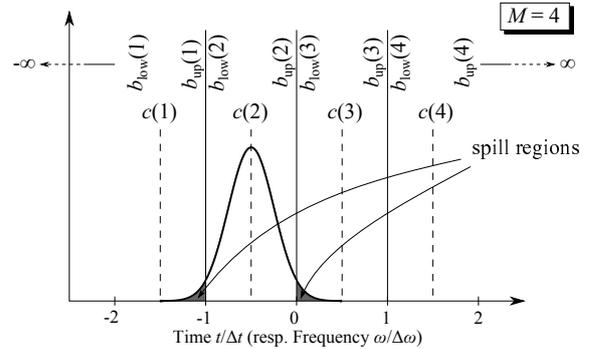

**Fig. 2: Definition of bins for $M = 4$.** Each symbol is associated to a symbol pulse located in a given bin. Bins, are defined in the time and frequency domains by their lower and upper bounds $b_{\text{low}}(j)$ and $b_{\text{up}}(j)$. The centers of the symbol pulses sent by the sender Alice are given by $c(i)$. As an example a pulse sent in the 2$^{\text{nd}}$ bin is shown. The spill regions, namely the regions which spill in other bins, are displayed in gray.

## 2.2 Mutual information

One tool for evaluating a QKD protocol is the mutual information. For one basis the mutual information of one qu-$N$-it in bits is

$$I'_{S,R} = \sum_{s=1}^{M}\sum_{r=1}^{M} P'_{S,R}(s,r) \log_2\left(\frac{P'_{S,R}(s,r)}{P'_S(s)P'_R(r)}\right), \quad (7)$$

where $M$ is the number of symbols, $s$ is a symbol out of the alphabet $S$ representing the sender and $r$ is a symbol out of the alphabet $R$ representing the receiver. $s$ out of $S$ is replaced by $a$ out of $A$ for Alice being the sender and $r$ out of $R$ is replaced by $b$ out of $B$ for the receiver being Bob and $e$ out of $E$ for it being Eve. $P'_{S,R}(s,r)$ is the joint probability of the sender sending the $s^{th}$ and the receiver receiving the $r^{th}$ symbol. $P'_S(s)$ (resp. $P'_R(r)$) is the probability of the sender sending (resp. the receiver receiving) the $s^{th}$ (resp. $r^{th}$) symbol. Up to $N = \log_2(M)$ bits per photon are possible. In this paper we want to calculate the mutual information considering that two bases are used with equal probability and with symbols 1 to $M$ being time symbols and $M+1$ to $2M$ being frequency symbols.

The formula for the mutual information needs to be modified, because only half of the $2M$ symbols can be used to create a secret key (only one basis is used at a time), so

$$\begin{aligned}I_{S,R} &= \sum_{s=1}^{2M}\sum_{r=1}^{2M} P_{S,R}(s,r) \log_2\left(\frac{P_{S,R}(s,r)}{P_S(s)P_R(r)}\right) - 1 \\ &= \sum_{s=1}^{2M}\sum_{r=1}^{2M} P_{R|S}(r|s) P_S(s) \log_2\left(\frac{P_{R|S}(r|s)}{P_R(r)}\right) - 1,\end{aligned} \quad (8)$$

which still results in up to $N = \log_2(2M) - 1 = \log_2(M)$ bits of information and can be rewritten with the relation $P_{S,R}(s,r) = P_{R|S}(r|s)P_S(s)$, where $P_{R|S}(r|s)$ is the conditional probability.

$\mathbf{P}_S$ and $\mathbf{P}_R$ are vectors with the $2M$ entries $P_S(s)$ resp. $P_R(r)$. Analogously $P_{R|S}(r|s)$ is an entry in column $s$ and row $r$ of the $2M \times 2M$ matrix $\mathbf{P}_{R|S}$ representing the transmission from the sender to receiver. $\mathbf{P}_{R|S}$ can be interpreted as follows: the columns represent which symbol a receiver will measure depending on the sent symbol represented by the rows. The conditional-probability matrices $\mathbf{P}_{R|S}$ will be created in the following section. The probability of receiving the different symbols can be calculated with

$$\mathbf{P}_R = \mathbf{P}_{R|S}\mathbf{P}_S, \quad (9)$$

assuming that Alice sends every symbol of both bases with the same probability, i.e. $P_A(a) = 1/(2M)$ for every $a$, $P_B(b)$ and $P_E(e)$ can be calculated.

## 2.3 Conditional probability for Bob

Because of the sifting process Bob will always be in the correct basis with respect to Alice. We begin with assuming both being in the PPM basis and (as mentioned earlier) perfectly rectangular filters of width $\Delta t$ for Bob:

$$\begin{aligned}P_{R|A}^{\text{correct}}(r|a) &= \int_{(b_{\text{low}}(r)-c(a))\Delta t}^{(b_{\text{up}}(r)-c(a))\Delta t} \rho_{\sigma_t}(t)\,dt \\ &= \frac{\sqrt{2}}{\sigma_t} \int_{(b_{\text{low}}(r)-c(a))\Delta t}^{(b_{\text{up}}(r)-c(a))\Delta t} \phi\left(\sqrt{2}\frac{t}{\sigma_t}\right)dt,\end{aligned} \quad (10)$$

and with the substitution $t = \Delta t\, x$ one finds

$$\begin{aligned}P_{R|A}^{\text{correct}}(r|a) &= \frac{\sqrt{2}\Delta t}{\sigma_t} \int_{b_{\text{low}}(r)-c(a)}^{b_{\text{up}}(r)-c(a)} \phi\left(\sqrt{2}\frac{\Delta t\, x}{\sigma_t}\right)dx \\ &= \frac{\sqrt{2}}{\alpha} \int_{b_{\text{low}}(r)-c(a)}^{b_{\text{up}}(r)-c(a)} \phi\left(\sqrt{2}\frac{2x}{\alpha}\right)dx.\end{aligned} \quad (11)$$

With a similar substitution one finds the same for Alice and Bob being in the frequency domain. Thus considering both bases we get

$$\mathbf{P}_{B|A}^{\text{Bob}} = \begin{pmatrix} \mathbf{P}_{B|A}^{\text{correct}} & 0_{M,M} \\ 0_{M,M} & \mathbf{P}_{B|A}^{\text{correct}} \end{pmatrix} \quad (12)$$

where $0_{M,M}$ represents $M \times M$ zero-matrices.

## 2.4 Conditional probability for Eve

Eve uses rectangular filters of width $\Delta t$ in one basis where each filter output forwards the photons to a second set of rectangular filters of width $\Delta \omega$ in the other basis whose outputs forwards the photons to $M^2$ detectors. Eve measuring in the PPM basis first when Alice sent in the PPM basis is analogous to Alice sending to Bob described by (11). The filtering taking place in the wrong basis is not relevant here. Without loss of generality, we assume that Eve always first filters in the PPM basis. Thus Eve's and Alice's bases differ when Alice sends FSK pulses, yielding a conditional probability for Eve's detected symbols of

$$P_{E|A}^{\text{wrong}}(e|a) = \int_{b_{\text{low}}(e)\Delta t}^{b_{\text{up}}(e)\Delta t} \rho_{\sigma_\omega^{-1}}(t)dt$$
$$= \frac{\sqrt{2}}{\sigma_\omega^{-1}} \int_{b_{\text{low}}(e)\Delta t}^{b_{\text{up}}(e)\Delta t} \phi\left(\sqrt{2}\frac{t}{\sigma_\omega^{-1}}\right)dt. \quad (13)$$

With the substitution $t = \Delta t\, x$ one gets

$$P_{E|A}^{\text{wrong}}(e|a) = \frac{\sqrt{2}\Delta t}{\sigma_\omega^{-1}} \int_{b_{\text{low}}(e)}^{b_{\text{up}}(e)} \phi\left(\sqrt{2}\frac{\Delta t x}{\sigma_\omega^{-1}}\right)dx$$
$$= \frac{\sqrt{2}}{\beta M} \int_{b_{\text{low}}(e)}^{b_{\text{up}}(e)} \phi\left(\sqrt{2}\frac{2x}{\beta M}\right)dx. \quad (14)$$

After the first filter the pulses will be broadened but still centered at the same position. Eve wants to preserve as much information as possible in the frequency domain without losing information for her time measurement, thus rectangular filters of width $\Delta t$ are already the optimal choice. Applying the Fourier transform, noted $\mathcal{F}(\circ)(\omega)$, to conjugated FSK pulses truncated in the time domain by the first filter leads to a modified Gaussian function in the frequency domain described by

$$g_f(\omega) = \left|\mathcal{F}\left(\psi_{\sigma_\omega^{-1}}(t) H_f\left(\frac{t}{\Delta t}\right)\right)(\omega)\right|^2$$
$$= \frac{\sigma_\omega}{\sqrt{\pi}} \left|\int_{-\infty}^{+\infty} \phi\left(\frac{t}{\sigma_\omega^{-1}}\right) H_f\left(\frac{t}{\Delta t}\right) e^{-i\omega t} dt\right|^2 \quad (15)$$

with the rectangular filter function being

$$H_f\left(\frac{t}{\Delta t}\right) = \begin{cases} 1 & \text{for } b_{\text{low}}(f) < t/\Delta t < b_{\text{up}}(f) \\ 0 & \text{otherwise.} \end{cases} \quad (16)$$

With the substitution $t = x\sigma_\omega^{-1}$ and $\omega = w\sigma_\omega$ we get

$$g_f(w) = \frac{1}{\sqrt{\pi}}\left|\int_{-\infty}^{+\infty}\phi(x) H_f\left(\frac{M\beta x}{2}\right)e^{-iwx}dx\right|^2. \quad (17)$$

Consequently

$$P_{E|A}^{\text{2nd correct}}(e|a) = \sum_{f=1}^{M}\int_{(b_{\text{low}}(e)-c(a))\Delta\omega}^{(b_{\text{up}}(e)-c(a))\Delta\omega} g_f(\tilde{\omega})d\tilde{\omega}, \quad (18)$$

with the sum over all the truncated pulses going through all time-filter outputs. Eve will save the information received in the time and frequency domain until the sifting process after which she only keeps the information about the announced bases.

Consequently the conditional probability matrix for Eve after the sifting process is

$$\mathbf{P}_{E|A}^{\text{Eve}} = \begin{pmatrix} \mathbf{P}_{E|A}^{\text{correct}} & 0_{M,M} \\ 0_{M,M} & \mathbf{P}_{E|A}^{\text{2nd correct}} \end{pmatrix}. \quad (19)$$

## 3 Optimal pulse widths

### 3.1 Symbol pulses

In the assumed intercept/resend attack Eve measures a fraction $\varepsilon$ of the photons sent by Alice and then resends a photon for each measured one.

With the conditional probabilities calculated in Section 2.3 and 2.4 and the corresponding mean probabilities we can calculate $I_{A,B}$ and $I_{A,E}$ from (8). Alice sends every symbol with the same probability, thus $P_A(a) = 1/(2M)$ for $a$ out of $A$. Bob does not know which photons exactly Eve attacked on. Thus Bob's probability and conditional probability is the mean of Eve attacking and not attacking:

$$\overline{\mathbf{P}}_{B|A} = (1-\varepsilon)\mathbf{P}_{B|A}^{\text{no attack}} + \varepsilon\mathbf{P}_{B|A}^{\text{attack}}, \quad (20)$$

$$\overline{\mathbf{P}}_B = (1-\varepsilon)\mathbf{P}_B^{\text{no attack}} + \varepsilon\mathbf{P}_B^{\text{attack}}. \quad (21)$$

If Eve is not attacking $\mathbf{P}_{B|A}^{\text{no attack}} = \mathbf{P}_{B|A}$ and $\mathbf{P}_B^{\text{no attack}} = \mathbf{P}_{B|A}\mathbf{P}_A$, see (12). If Eve is attacking, she resends photon symbol depending on what she has measured. Eve gets information of both bases. We assume Eve sends in the basis in which she potentially has more information, hence in which she filters first (here in the PPM basis). Eve wants to mimic Alice, so she uses the same pulse relations as her.

If Eve attacks, there are two possibilities: Firstly, Eve is in the correct basis compared to Alice and subsequently resends a photon according to what she has measured. Thus the symbol probability matrix for Bob is $\left(\mathbf{P}_{B|A}^{\text{correct}}\right)^2$ (the square represents the matrix product with itself). Secondly, Eve is in the wrong basis compared to Alice, consequently Bob will be in the wrong basis compared to Eve. Thus the conditional probability for him is $\mathbf{P}_{B|A}^{\text{wrong basis}}$. Both

scenarios happen with equal probability, so the conditional probability of Bob if Eve attacks is

$$\mathbf{P}_{B|A}^{\text{attack}} = \frac{1}{2}\begin{pmatrix} \mathbf{P}_{B|A}^{\text{mean}} & 0_{M,M} \\ 0_{M,M} & \mathbf{P}_{B|A}^{\text{mean}} \end{pmatrix} \quad (22)$$

with $\mathbf{P}_{B|A}^{\text{mean}} = \left(\mathbf{P}_{B|A}^{\text{correct}}\right)^2 + \mathbf{P}_{B|A}^{\text{wrong}}$. Furthermore we get $\mathbf{P}_B^{\text{attack}} = \mathbf{P}_{B|A}^{\text{attack}} \mathbf{P}_A$ for Bob's measurement outcome. With (8) we can now calculate the mutual information of Alice and Bob:

$$I_{A,B} = \sum_{b=1}^{2M}\sum_{a=1}^{2M} \overline{P}_{B|A}(b|a) \\ \times P_A(a) \log_2\left(\frac{\overline{P}_{B|A}(b|a)}{\overline{P}_B(b)}\right) - 1. \quad (23)$$

Eve's measurement outcome is

$$\mathbf{P}_E = \mathbf{P}_{E|A}^{\text{Eve}} \mathbf{P}_A. \quad (24)$$

She only gets information when she is eavesdropping, thus the mutual information between Alice and Eve is

$$I_{A,E} = \varepsilon\left[\sum_{e=1}^{2M}\sum_{a=1}^{2M} P_{E|A}^{\text{Eve}}(e|a) P_A(a) \\ \times \log_2\left(\frac{P_{E|A}^{\text{Eve}}(e|a)}{P_E(e)}\right) - 1\right]. \quad (25)$$

The secret capacity, thus the upper bound for the number of secret bits per photon, for infinite keys (see for example [18]) is

$$C = \begin{cases} I_{A,B} - I_{A,E} & \text{for } I_{A,B} > I_{A,E} \\ 0 & \text{for } I_{A,B} \leq I_{A,E} \end{cases}. \quad (26)$$

Note that for $M = 2$ the secret capacity is the secret fraction as used for example in [18]. $C$ can become larger than one for $M > 2$ and represents the secret bits per photon, thus is rather a capacity than a fraction of bits.[1]

The parameters of our protocol should be chosen such that $C$ is maximized. Fig. 3 provides an optimization as a function of $\alpha$ and $\beta$. $C$ strongly depends on $\alpha$ but hardly on $\beta$ (for $\alpha$ being near the optimum).

---

[1] The secret key rate $S = KC$ of a realization of the protocol can be derived from the sifted key rate $K$ and $C$ depending on $\alpha$ and $\beta$ and matching QSER $= 1 - 1/(2M)\,\text{trace}\left(\overline{\mathbf{P}}_{B|A}\right)$. For the correlation between QSER and quantum bit error rate see [19].

In the following section we will see how to still find an optimal value of $\beta$ for more general eavesdropping strategies.

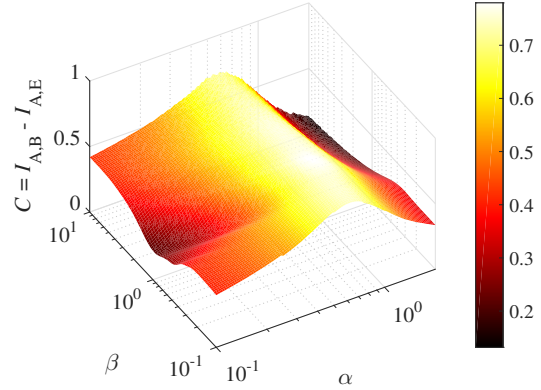

**Fig. 3:** (Color online) Secret capacity $C$ for $M = 4$ as a function of the normalized pulse width $\alpha$ and conjugate pulse width $\beta$.

## 3.2 Conjugated pulses

In the described intercept/resend attack Eve does not exploit the imperfect overlap of time and frequency symbols, i.e. the partial available information on the used basis. One example of such a strategy is described in [15]. Such an attack benefits from the capability of Eve to perform a basis-dependent attack which is only possible if she can discriminate the bases. This is mainly depending on the conjugated-pulse overlap with the symbol pulses.

In [12] the security of CV-TF-QKD is shown for a certain pulse relation by showing that the equality

$$\sum \rho_{\sigma_t}(t) = \sum \rho_{\sigma_\omega}(t) \quad (27)$$

holds for the sum over all possible pulses. In the DV case this equation cannot be fulfilled. Our approach is to quantify the deviation from (27) by evaluating the difference $\sum \rho_{\sigma_t}(t) - \sum \rho_{\sigma_\omega}(t)$, which in our case is not equal to zero. Considering all pulses, the difference can be written as

$$U_\alpha(\beta) \equiv \int \sum_{s=1}^{M} \left| \rho_{\sigma_t}\left(t + \left[s - \frac{M+1}{2}\right]\Delta t\right) \right. \quad (28)$$
$$\left. - \rho_{\sigma_\omega^{-1}}(t) \right| dt.$$

$U_\alpha(\beta)$ can then be minimized numerically in order to find and optimum normalized conjugated pulse width $\beta_{\text{opt}}$.

### 3.3 Secret key rate

In the following, the results of maximizing (26) and subsequently minimizing (28) are presented and discussed.

$C$, $\alpha_{\text{opt}}$ and $\beta_{\text{opt}}$ are plotted over $M$ in the top of Fig. 4 for different $\varepsilon$. $\alpha_{\text{opt}}$ approaches a value between 0.4 and 0.6 and $\beta_{\text{opt}}$ a value of roughly 0.7 for a higher number $M$ of symbols per basis. In the bottom of Fig. 4 the secret capacity $C$ is plotted as a function of $M$. While $C$ becomes smaller with a higher fraction $\varepsilon$ of eavesdropped photons it increases with a higher $M$. For $\varepsilon \geq 0.9$ the capacity vanishes for all $M$. It is obvious that, when no eavesdropper is present, the symbol pulses should be as narrow as possible to prevent spill-region induced bit errors. With a higher eavesdropping fraction $\varepsilon$ the optimum symbol pulses get wider to prevent Eve from obtaining information. The conjugated pulses follow the symbol pulses in width to increase the overlap.

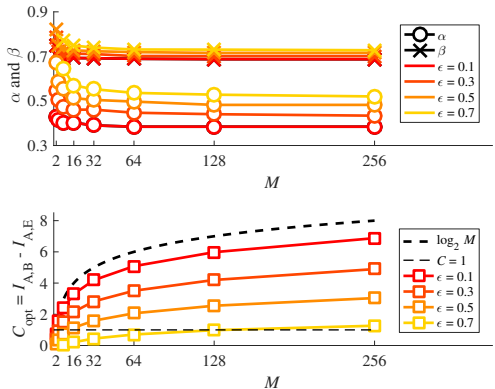

**Fig. 4:** (Color online) Top: $\alpha_{\text{opt}}$ and $\beta_{\text{opt}}$. Bottom: the secret capacity (secret bits per photon) $C$ for different $M$ and $\varepsilon$ calculated with the optimal pulse widths $\alpha_{\text{opt}}$ and $\beta_{\text{opt}}$.

If $\varepsilon$ approaches zero, $\alpha$ gets small which leads to $C \approx \log_2 M$. Comparing that to QKD protocols like BB84 ($C \approx 1$ for low errors), TF-QKD can have a tremendous advantage for repetition rates much smaller than $1/\Delta t$. For a 100 MHz repetition rate $M = 256$ PPM-symbols fit ($\Delta t \approx 39$ ps) which leads to up to eight times higher secret key rates (See dashed lines in Fig. 4, assuming $M = 256$ FSK-symbols).

### 4 Conclusion

In this paper we provided a detailed analysis of the DV-TF-QKD protocol. We show that symbols with infinitesimal width are not optimal, quite contrarily a certain width is preferable, which contradicts [15].

We have discussed that a complete optimization of an even slightly more advanced protocol against all possible attacks under realistic conditions is very challenging. Our analysis suggests that efforts on full numerical assessment of QKD protocols are rewarding.

By a numerical approach assuming an intercept/resend attack and a high overlap between pulses of both bases the optimal pulse widths were calculated. Although a security proof against general attacks is still missing, these pulse widths can be used as a good estimation on how to form the bases in a real implementation of DV-TF-QKD. Additionally the secret bits per photon were calculated for different parameters under the assumptions above. Further it was shown that a higher number of symbols per basis increases the secret key rate.

Extending the implementation for $M = 2$ [13] to higher $M$ would be possible by cascading the used time and frequency filters or by using detectors which can measure time respective frequency directly.

The DV-TF-QKD protocol is a promising QKD protocol, especially regarding the high number of possible bits per photon for lower repetition-rates. Compared to its CV counterpart it is easier to implement mostly using off-the-shelf telecom components. PPM and FSK rely on standard techniques in classical communication, which makes this protocol well suited for both free-space and fiber based QKD.

This work has been funded by the German Research Foundation (DFG) within the CRC 787, project C2.